\documentclass[letterpaper]{article} % DO NOT CHANGE THIS
\usepackage{aaai23}  % DO NOT CHANGE THIS
\usepackage{times}  % DO NOT CHANGE THIS
\usepackage{helvet}  % DO NOT CHANGE THIS
\usepackage{courier}  % DO NOT CHANGE THIS
\usepackage[hyphens]{url}  % DO NOT CHANGE THIS
\usepackage{graphicx} % DO NOT CHANGE THIS
\usepackage{natbib}  % DO NOT CHANGE THIS AND DO NOT ADD ANY OPTIONS TO IT
\usepackage{caption} % DO NOT CHANGE THIS AND DO NOT ADD ANY OPTIONS TO IT
\usepackage{algorithm}
\usepackage{algorithmic}
\usepackage{amsmath}
\usepackage{amssymb}
\usepackage{multirow}
\usepackage{subfigure}
\usepackage{multicol}
\usepackage{makecell}
\usepackage{booktabs}
\usepackage{newfloat}
\usepackage{listings}
\DeclareCaptionStyle{ruled}{labelfont=normalfont,labelsep=colon,strut=off} % DO NOT CHANGE THIS
\floatstyle{ruled}
\newfloat{listing}{tb}{lst}{}
\floatname{listing}{Listing}
\title{Similarity Distribution based Membership Inference Attack on Person Re-identification}
\author{
    Junyao Gao\textsuperscript{\rm 1}, Xinyang Jiang\textsuperscript{\rm 2}, Huishuai Zhang\textsuperscript{\rm 2}, Yifan Yang\textsuperscript{\rm 2}, Shuguang Dou\textsuperscript{\rm 1},\\
    Dongsheng Li\textsuperscript{\rm 2}, Duoqian Miao\textsuperscript{\rm 1}, Cheng Deng\textsuperscript{\rm 3}, Cairong Zhao\equalcontrib\textsuperscript{\rm 1}\\
}
\affiliations{
    \textsuperscript{\rm 1}Tongji University\\
    \textsuperscript{\rm 2}Microsoft Research Asia\\
    \textsuperscript{\rm 3}Xidian University
}

\usepackage{bibentry}
\begin{document}

\maketitle

\begin{abstract}
While person Re-identification (Re-ID) has progressed rapidly due to its wide real-world applications, it also causes severe risks of leaking personal information from training data. Thus, this paper focuses on quantifying this risk by membership inference (MI) attack. Most of the existing MI attack algorithms focus on classification models, while Re-ID follows a totally different training and inference paradigm. Re-ID is a fine-grained recognition task with complex feature embedding, and model outputs commonly used by existing MI like logits and losses are not accessible during inference. Since Re-ID focuses on modelling the relative relationship between image pairs instead of individual semantics, we conduct a formal and empirical analysis which validates that the distribution shift of the inter-sample similarity between training and test set is a critical criterion for Re-ID membership inference. As a result, we propose a novel membership inference attack method based on the inter-sample similarity distribution. Specifically, a set of anchor images are sampled to represent the similarity distribution conditioned on a target image, and a neural network with a novel anchor selection module is proposed to predict the membership of the target image. Our experiments validate the effectiveness of the proposed approach on both the Re-ID task and conventional classification task.

\end{abstract}

\section{Introduction}
Nowadays, the deep learning model has made remarkable progress with wide applications but also exposes risks of leaking personal information from its training set \cite{fredrikson2015model,wu2016methodology,shokri2017membership}, especially for tasks where the data source is sensitive like person re-identification (Re-ID). 
Re-ID is an image retrieval task that  identifies a specific person in different images or video sequence scenes. 
A Re-ID training set contains pedestrian images, and  leaking information from it causes serious social security and ethical risks. 
For example, the attacker could access the movements of a particular person in different places to conspire some evil plots. 
Therefore, it becomes necessary to quantify the information leakage of Re-ID data.

One common methodology to quantify the privacy risk of a trained model is using the attack success rate of membership inference (MI)  attack \cite{shokri2017membership,yeom2018privacy,salem2018ml,long2018understanding,nasr2018machine,song2019privacy,chen2021machine}. MI attack algorithm infers whether a record belongs to the training set by some information of target model and is generally described as a binary classification problem.

\begin{figure*}
    \centering
    \includegraphics[width=2.1\columnwidth]{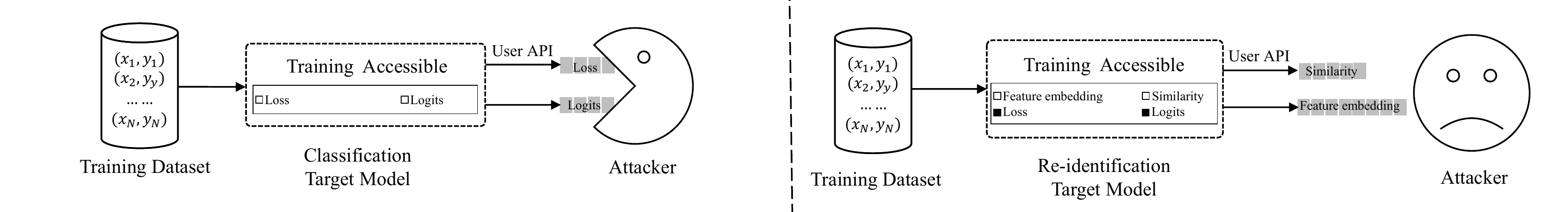}
    \caption{The different outputs for  classification model and Re-ID model under the black-box setting. For classification model (left), adversarial   can access the logits  and loss both during and after the training processing. However, for the Re-ID model (right), only  similarity and feature embedding are accessible during inference, which is not suitable for existing classification-based MI attacks. 
    }
    \label{fig1}
\end{figure*}

Most of existing MI attack methods focus on classification task \cite{shokri2017membership,yeom2018privacy,sablayrolles2019white}, where the attacker infers the membership of a sample based on its corresponding  model outputs, % highly related to the generalization gap
 such as logits (i.e. prediction confidence of each category) or loss (inferred from logits and ground-truth label), as shown in Figure \ref{fig1}.

However, compared to classification, Re-ID follows a totally different training and inference paradigm, bringing new challenges to existing MI attack methods. 
State-of-the-art (SOTA) Re-ID methods first  extract visual features from each pedestrian image and then conduct recognition by retrieving images based on the relative similarity between image pairs. 
During training, SOTA Re-ID methods add an extra identity classifier after the feature extractor, which are not available during inference. 
As a result, the attacker generally only gets the feature embedding of individual images,  while the commonly used logits or loss for MI attack on classification are not available in the Re-ID task.
As validated by previous works \cite{nasr2018comprehensive}, compared to logits and loss, feature embedding contains more information irrelevant to training data and does not characterize the training-test generalization gap well. 
Furthermore, compared to the general classification, Re-ID is a more challenging fine-grained recognition task, leading to a more complex and less discriminative feature distribution for MI attacks. 

As a result, in this paper, instead of the conventional model outputs like feature, logits and loss, we aim at finding a new set of features specifically for MI attack on person re-identification. 
Compared to classification tasks that focus on the semantics of individual samples, Re-ID is a metric learning task  modelling the relative relationship between image pairs. 
Thus, instead of looking at the features of individual images, we tackle these challenges by elaborately exploring the inter-sample correlation between different images and studying how the generalization gap of the Re-ID model affects the distribution of pair-wise similarity. 
Intuitively, the Re-ID model explicitly pushes  images with the same identities on the training set together and pulls ones with different identities away from each other \cite{oh2016deep,duan2017deep,ming2022deep}, which will be difficult to fully generalize to samples in the test set, resulting in an inter-image similarity distribution shift between training and test set.
This intuition is validated by our formal analysis of optimal attack as well as the preliminary experiments in section 3. 
Our experiments compare the statistical properties of the inter-sample similarity distribution of training samples and test samples, showing an obvious difference between training and test set. 

Based on the analysis, we propose a novel MI attack method called \textit{similarity distribution based MI attack} (SD-MI attack), which conducts membership inference by exploiting the relative correlation between image pairs. 
Specifically, given a target image, the inter-sample similarity distribution conditioned on the target image is represented by a set of sampled anchor images, and the membership of the target image is inferred based on its similarity with the anchors within the reference set by a neural network.
In order to select appropriate anchor images that better represent the similarity distribution, we propose to use an attention-based neural module that is able to automatically select anchor images based on their feature embedding.
Extensive experiments demonstrate that our approach outperforms existing MI attack algorithms on general Re-ID models. 

The contributions of our work are summarized as follows.
1) We raise a rarely studied privacy risk of the training set in the Re-ID task, whose information leakage is quantified by our proposed MI attack algorithm; 
2) We propose the first MI attack algorithm on person Re-identification, which attacks a target image by exploiting its relative correlation with reference images;
3) The proposed method outperforms existing MI attack approaches on Re-ID models. 
We hope this work will attract more  attention to the data privacy risk of person re-identification, as well as more and more computer vision tasks other than classification.

\section{Related Work}
Because of the high complexity of the training set and the target model, it is extremely difficult to theoretically analyze  why membership inference attacks work. Recent work by \cite{yeom2018privacy} describes the generalization gap of the target model as the main reason that affects membership inference attack success rate. \citet{shokri2017membership,sablayrolles2019white} observe that the attack model is more likely to infer membership when the target model performs better on the training set than on the test set.  \citet{Li2020MembershipIA} experimentally demonstrate that an upper bound of membership inference attack success rate is determined by the generalization gap of target model. Most MI attack issues are essentially based on the prediction vectors forthrightly related to the generalization gap, such as loss and logits. However, our paper makes a thorough inquiry on the distribution gap of similarities. 

\citet{li2022user} propose a user-level MI attack in metric embedding learning. This approach is based on an assumption that data from the same category forms a more compact cluster in the training set than the test set,  and uses the average and pair-wise intra-class distance as features to  conduct user-level membership inference. However, this method requires multiple samples from one class and the number of sample in the class severely affect the attack success rate (i.e. low number of samples causes low attack success rate). 
While \citet{li2022user} only focuses on average and pair-wise distance on intra-class samples, our method proposes to look at more general similarity distribution over all sample pairs (both intra- and inter-class similarity). Furthermore, our method does not require multiple samples for each identity. 

\citet{shokri2017membership} firstly propose the approach that trains a binary classifier to conduct membership inference on classification model by using logits as features. For the generative adversarial network, \cite{hayes2017logan} believe that trained generator will lead the stronger confidence scores on the training set. In this paper, we establish \textit{the similarity distribution} membership inference attack approach that describes the distribution gap between training and test set of trained Re-ID model by the similarity with the target image $x_t$ and the anchor images. 

\section{Preliminary Analysis}
\subsection{Preliminaries}
In this paper, we focus on the most effective type of approaches adopted by most of the existing state-of-the-art Re-ID models, which use a softmax-based classifier as a loss function. 
Given a Re-ID dataset $D$ containing images sampling from a data distribution $P(x)$ as the form of $( x , y) \in {\mathcal{X}} \times \mathcal{Y}$, where ${x}$  is the pedestrian image and $y$ is the identity label corresponding to ${x}$. 
Existing methods \cite{zheng2016person,hu2017person} consider Re-ID as an image classification task during training, where an image ${x}$ is fed into a backbone network to extract  high-dimensional features, which is then fed into fully connected layers to classify ${x}$ with an identity $y$. 
The cross-entropy loss is applied to train the classification model:

\begin{equation}
    \mathcal{L}_{id}=\frac{1}{n}\sum_{i=1}^{n}\log(p(y_{i}|x_{i})) \label{formula1}
\end{equation}

%Specifically, an identity-classification model is first trained on a private train dataset during training. 
During inference, given a query image, Re-ID is essentially an image retrieval task where the goal is to find the images with the same identity as the query image from a gallery. 
This is achieved by removing the identity classifier and using the high level feature before the classifier  to compute the similarity between the query image and gallery image. Then, person re-identification is conducted by sorting images based on this similarity. 

\subsection{Optimal Membership Inference}
We follow the assumption in \cite{sablayrolles2019white} that models the posterior distribution of model parameters $\theta$ as:
\begin{equation}
\small
    P(\theta|\{(x_i, y_i, m_i)\}_{i=1}^n) \propto \exp\left(-\frac{1}{T}\sum_{i=1}^n m_i\mathcal{L}(\theta, x_i, y_i)\right)
    \label{eq_parameter_posterior}
\end{equation}
where $m_i$ is the membership variable for each sample that $m_i=0$ means test set and $m_i=1$ means training set. And $T$ is a temperature parameter  controlling the stochasticity of $\theta$.
Substituting the Re-ID loss function into Eq.\ref{eq_parameter_posterior}, the posterior distribution of Re-ID model parameters is:
\begin{equation}
\small
\begin{aligned}
  P(\theta|\{(x_i, y_i, m_i)\}_{i=1}^n) & \propto \exp\left(-\frac{1}{T}\sum_{i=1}^n m_i\mathcal{L}(\theta, x_i, y_i)\right) \\
  & = \exp\left(-\frac{1}{T}\sum_{i=1}^n m_i \log P(y_i|x_i;\theta) \right) \\
  & = \exp\left(-\frac{1}{T}\sum_{i=1}^n m_i \log \frac{d(x_i, a_{y_i})}{\sum_{j=1}^kd(x_i, a_j)} \right)
    \label{eq_parameter_posterior_reid}
\end{aligned}
\end{equation}
where $d(x_i, a_j)$ denotes a similarity measurement in Re-ID representation space, which has multiple variants for different cross-entropy based Re-ID methods, such as L2Softmax \cite{ranjan2017l2} and AngularSoftmax \cite{10.5555/3045390.3045445},  $a_j$ is a learned class centers  representing each identities and $k$ is the number of identities. 

Following \cite{sablayrolles2019white}, given the set of other samples and their membership $\mathcal{T}=\{(x_i, y_i, m_i)\}_{i=1}^n$, the membership of the sample $x_1$ is inferred as:
\begin{equation}
\small
\begin{aligned}
&\mathcal{M}(\theta, x_1, y_1):=P(m_1=1|\theta, x_1, y_1) \\
&=E_{\mathcal{T}}
\left[
    \sigma\left(
        s(x_1, y_1, \theta, P(\theta|\mathcal{T}))+\log\frac{P(m_1 = 1)}{1-P(m_1 = 1)}
    \right)
\right]
\end{aligned}
\label{eq_mia}
\end{equation}
where
\begin{equation}
\small
\begin{aligned}
&s(x_1, y_1, \theta, P(\theta|\mathcal{T})) =- \frac{1}{T}\log \frac{d(x_1, a_{y_1})}{\sum_{j=1}^kd(x_1, a_j)} \\
&-\log \left(
\int_{\theta'}
\exp\left(
-\frac{1}{T}\log \frac{d(x_1, a_{y_1})}{\sum_{j=1}^kd(x_1, a_j)}P(\theta'|\mathcal{T}) 
\right)
d\theta' 
\right)
\end{aligned}
\label{eq_mia_s}
\end{equation}

As shown in Eq.\ref{eq_mia} and Eq.\ref{eq_mia_s}, the  second term of Eq.\ref{eq_mia_s} corresponds to the typical loss of $x_1$ under the models that have not trained with $x_1$ and can be seen as a threshold for MI attack. If this term is computed or properly approximated, the optimal membership inference depends only on the relative similarity between target sample $x_i$ and the identity centers $a_j$. 
However, as discussed in the introduction, these learn-able identity centers are usually not accessible for attackers. 
As a result, since the Re-ID loss pushes the training samples to their corresponding centers as close as possible, it is intuitive to select a set of proxy centers to approximate the learned centers (called anchor images in this paper) from the actual Re-ID dataset images and conduct membership inference based on the sampled proxy centers. 
Our preliminary experiments in the next sub-section verify that there is an obvious and distinguishable difference between the statistical properties of the similarity between the target image and randomly sampled anchor images in  the training and test set.  

\subsection{Preliminary Experiments}

\begin{figure}
\centering
\subfigure[Average of similarities]{\includegraphics[width=0.47\columnwidth]{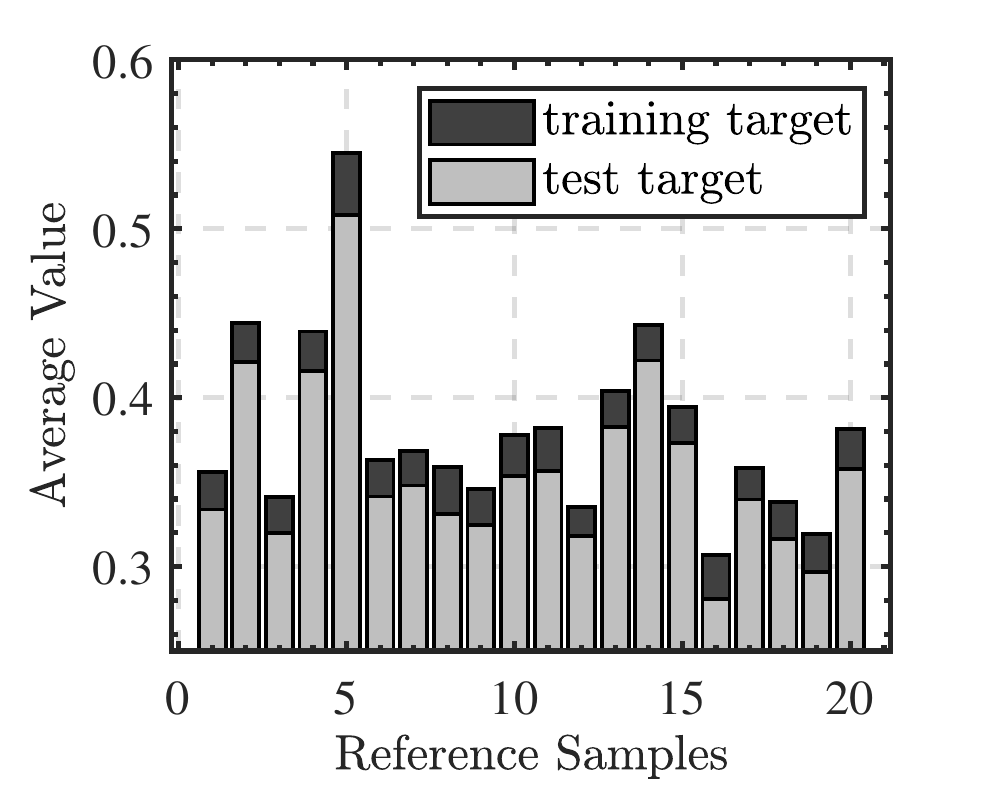}}
\subfigure[Std of similarities]{\includegraphics[width=0.47\columnwidth]{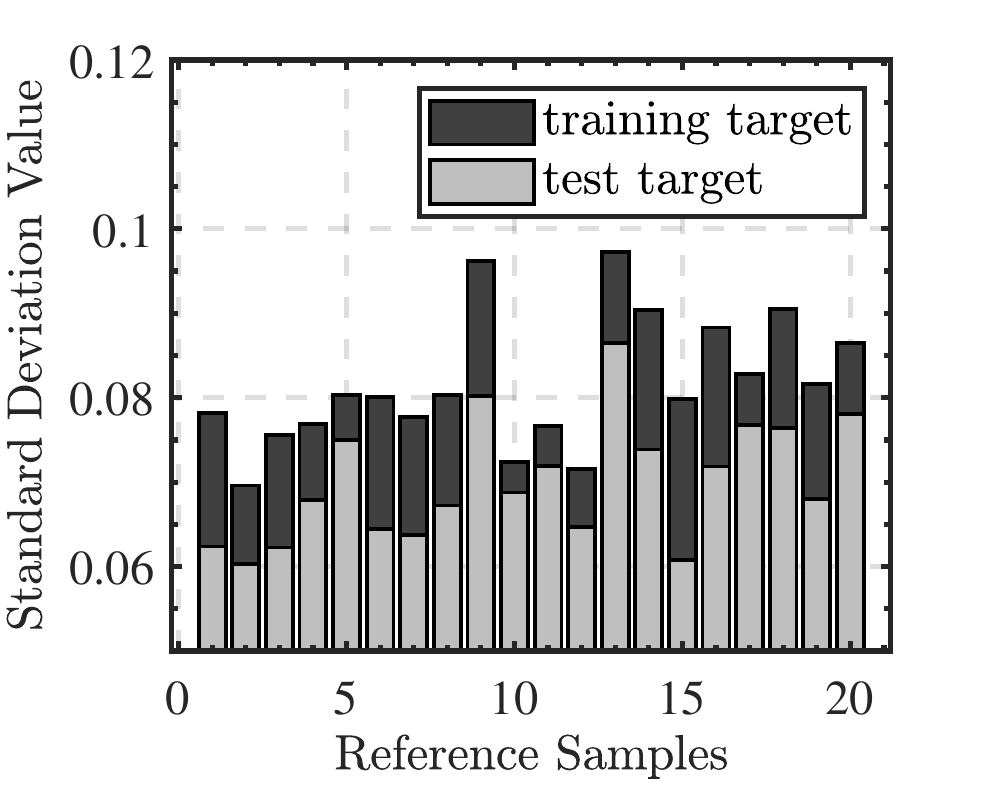}}
\caption{The average and standard deviation gap of distance from every reference sample to training target images or test target images.} \label{fig2}
\end{figure}

\begin{figure}
\centering
\subfigure[Average of similarities]{\includegraphics[width=0.47\columnwidth]{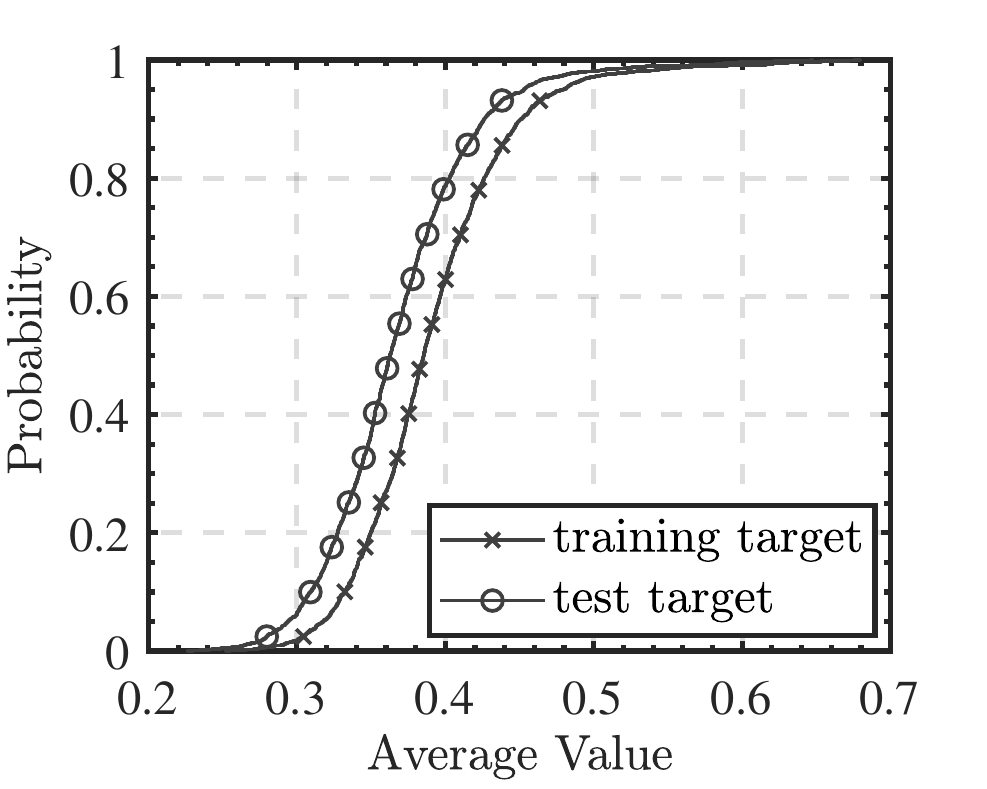}}
\subfigure[Std of similarities]{\includegraphics[width=0.47\columnwidth]{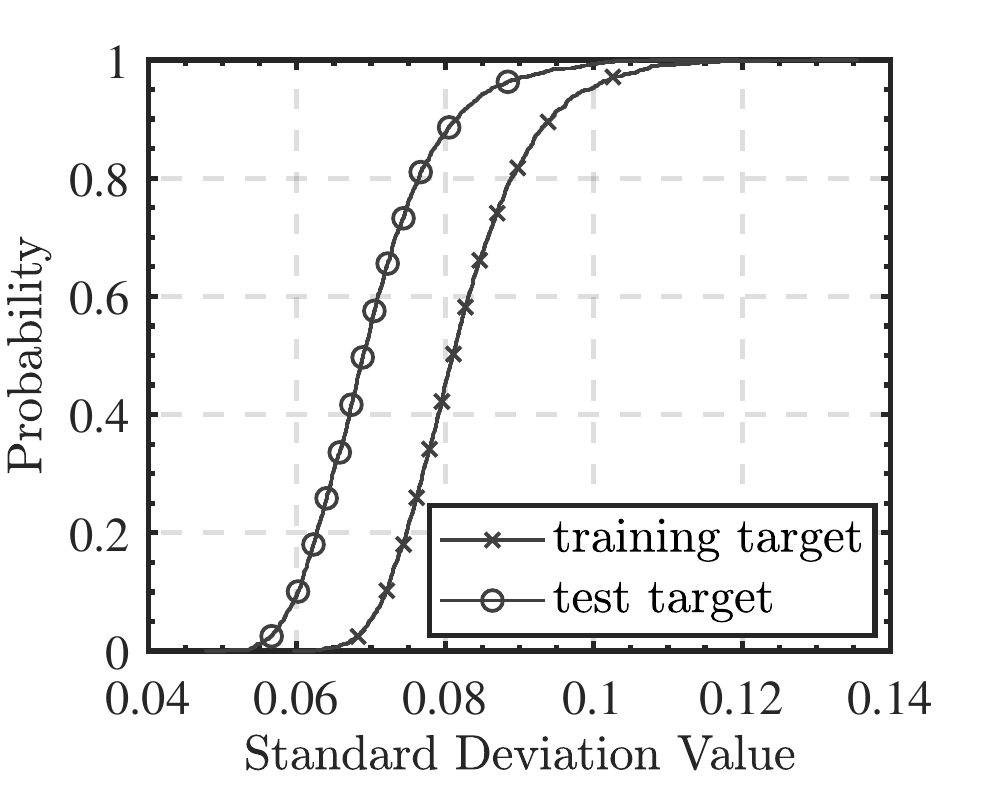}}
\caption{Cumulative density function of the average and standard deviation of the distance from the all reference samples to training target images and test target images.} 
\label{fig3}
\end{figure}

\begin{figure*}
    \centering
    \includegraphics[width=2.0\columnwidth]{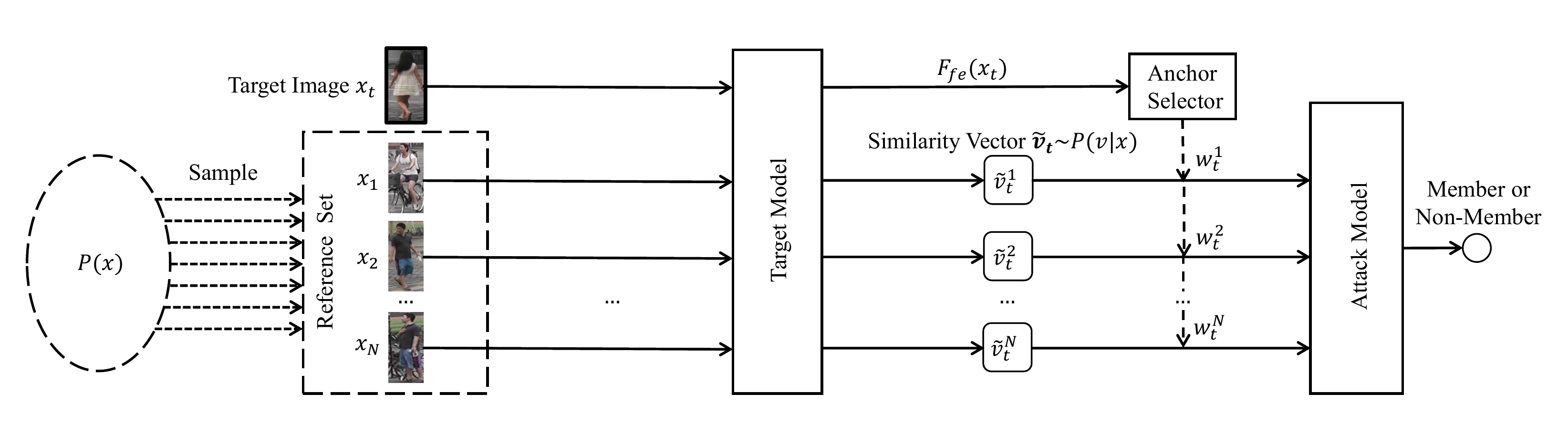}
    \caption{The two-stages pipeline of our black-box MI attack. First, for each target image $x_t$ we compute the similarity vector $\boldsymbol{\tilde{v}_t}$ with reference samples. Second, we fed similarity vector $\boldsymbol{\tilde{v}_t}$ into the attack model to infer the membership of target image $x_t$. Furthermore, we propose the anchor selector module  selecting useful  anchor images in the limited reference set to better approximate the similarity distribution. }
    \label{fig4}
\end{figure*}

\noindent\textbf{Experiment Configuration.} The formal analysis in the last sub-section has shown that the membership of a target image depends on the relative similarity between the target image and the identity centers learned from Re-ID training. 
Since identity centers are not accessible for MI attacks, we propose to sample a set of reference images from the Re-ID dataset as proxy centers and examine how the training/test generalization gap affects  their similarities with the target image. 
Specifically, given a dataset containing training samples denoted as $D_{train}$  and another dataset containing test samples denoted as $D_{test}$, we sample an extra set of 10\% reference samples and obtain the Euclidean distance between target samples in $D_{train}$/$D_{test}$ and the  reference samples. 

\noindent\textbf{Statistical Analysis.} 
From the distance matrix, we observe that the individual pair-wise distance has a high standard deviation and does not show obvious patterns relating to membership. 
Thus, several statistical properties of the overall distance distribution are compared between the distances from a training sample and a test sample to the reference samples.  

Firstly, we examine the mean of the distance from each target sample in $D_{train}$ / $D_{test}$ to different reference samples, as shown in Figure \ref{fig2} (a). 
The y-axis corresponds to the average distance from each target sample in $D_{train}$ or $D_{test}$ to a specific reference sample, and the x-axis refers to different reference samples. 
As a result, we observe a clear margin between the mean distance corresponding to  $D_{train}$ and the mean distance corresponding to $D_{test}$. The average distance corresponding to target samples in $D_{train}$  is generally larger than samples in $D_{test}$. 
Similarly, the standard deviation of the distance from each target sample in $D_{train}$ / $D_{test}$ to different reference samples is shown in Figure \ref{fig2} (b), which also shows a clear margin between samples from $D_{train}$ and samples from ${D_{test}}$. 

Besides looking at the mean and deviation of distance based on each individual reference image, we further examine the distribution of the mean and standard deviation over all reference images, which is represented as a cumulative distribution function as shown in Figure \ref{fig3}.  We observe that the cumulative distribution functions corresponding to samples in $D_{test}$ are always above those corresponding to the samples in $D_{train}$.

\noindent\textbf{Design Principles.} 
In conclusion, our experiments show that there is an obvious similarity distribution shift between training and test set, which means the similarity distribution between the target sample and a set of anchor images is an effective feature for membership inference. 

\section{Proposed Method}
We first briefly introduce the overall pipeline of our Similarity Distribution based Membership Inference Attack, as shown in Figure \ref{fig4}. 
Our method mainly contains two stages. In the first stage, given a target image, we obtained a feature vector that represents the conditional  distribution of the similarity between the target images and other images in the data distribution.
In the second stage, the membership inference is conducted based on the similarity distribution with a novel neural network structure. 
In the next two sub-section, we will elaborate on our designs and implementation in the two stages respectively. 

\subsection{Obtaining Similarity Distribution}
Following the design principles, the membership of a target image $x_t$ is inferred based on its similarity with a set of anchors sampled from the Re-ID data distribution $P(x_t)$. 

Specifically, we first sample a set of anchor images from the Re-ID data distribution $P(x)$, i.e. a reference set $\boldsymbol{r_t} = [r^1_t, r^2_t, r^3_t, \dots , r^N_t]$ where $r^i_t \in D$ is randomly sampled from dataset $D$ and $N$ is the image number of reference set. The $i$-th sampled distance $\tilde{v}^i_t$ of sampled similarity vector $\boldsymbol{\tilde{v}_t} = [\tilde{v}^1_t, \tilde{v}^2_t, \tilde{v}^3_t, \dots , \tilde{v}^N_t]$ is obtained by computing the euclidean distance between the target image $x_t$ and the $i$-th anchor image $r^i_t$ in  reference set:
\begin{equation}
        {\tilde{v}^i_t}  = {\Vert F_{fe}(x_t) - F_{fe}( r^i_t) \Vert}_2^2, %\ \tilde{v}_i \sim P(\widetilde{V}|X = x)
\end{equation}
where $F_{fe}$ is a function that map any input instance to its feature embedding in the target model. Concretely, we assume the feature embedding $F_{fe}(x_t)$ and $F_{fe}( r^i_t)$ for the target image $x_t$ and the reference image $r^i_t$ are the points in $K$-dimensional Euclidean space. 
The sampled similarities are then constructed as a similarity vector $\boldsymbol{\tilde{v}_t}$, which is then fed into our membership inference network to predict the membership of the target sample $x_t$.

\begin{figure}[ht]
    \centering
    \includegraphics[width=0.9\columnwidth]{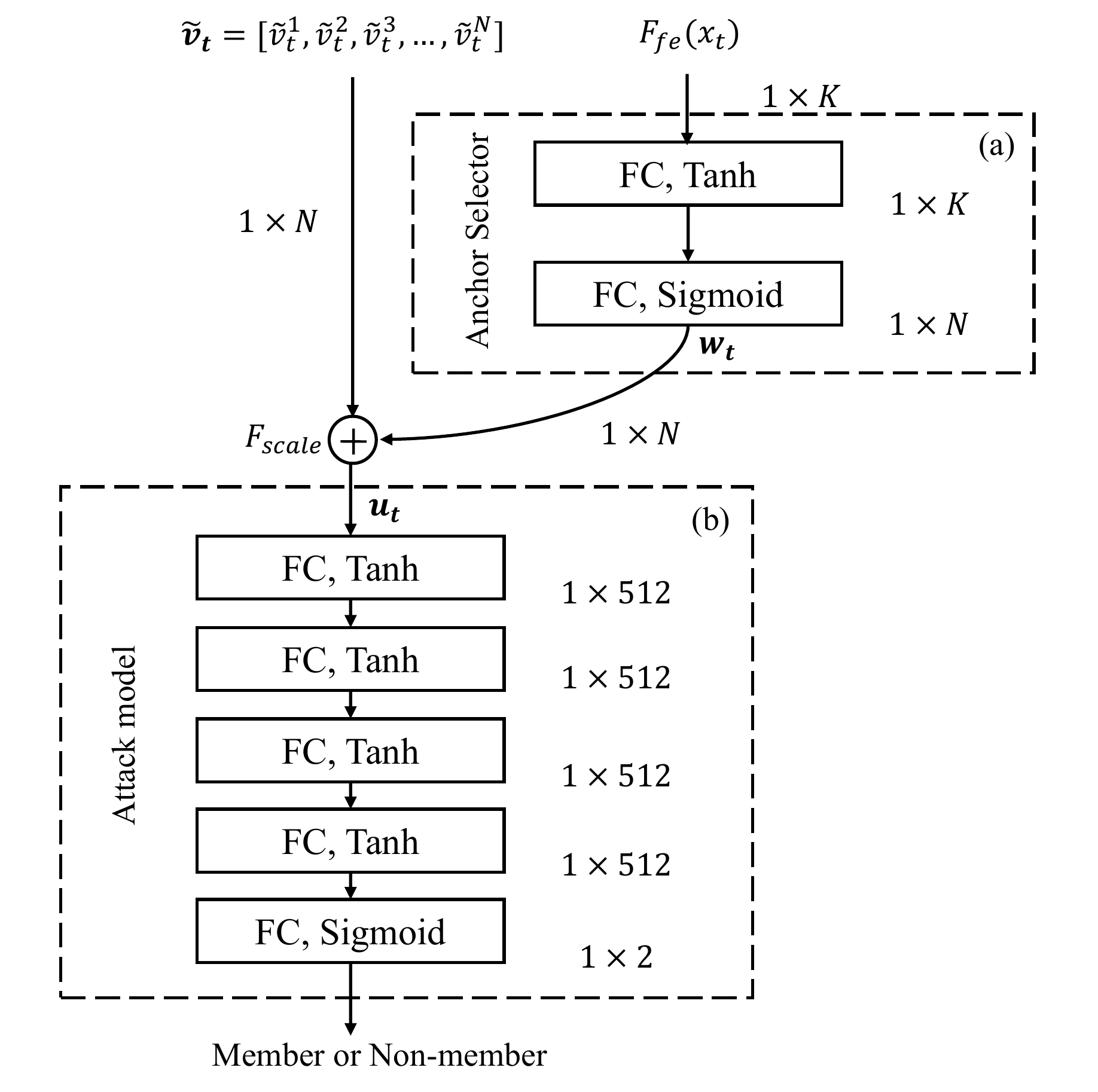}
    \caption{The specific architectures of our attack model (b) and  anchor selector module (a).}
    \label{fig5}
\end{figure}

\subsection{Membership Inference Network}
Figure \ref{fig5} (b) is a model structure of our proposed membership inference network. It takes the similarity vector between a target image and a reference set of anchor images as input and outputs a binary value to decide the membership. 

Following previous works \cite{shokri2017membership,long2018understanding,salem2018ml,yu2021does,chen2021machine}, we apply a multi-layer perceptron (MLP)  for membership inference. We build four hidden layers with 512 neurons and Tanh non-linearity activation and a binary classification output layer with sigmoid activation as our attack model for SD-MI attack. We refer to this approach as $M_{SD}$.
As discussed in the previous section, Eq.\ref{eq_mia_s} shows that selecting appropriate reference images to better approximate the identity centers $a_j$ is essential to Re-ID membership inference. 
As a result, we add an extra \textit{anchor selector module} to select appropriate anchor images with regards to the current image content, which assigns  weights $w^i$ to the distances to different reference images.
As shown in Figure \ref{fig5} (a), the anchor selector $F_{as}$ takes the high-dimensional Re-ID feature embedding $F_{fe}(x_t)$ of the target image $x_t$ as input.  
We implement this module $F_{as}$ as a 2-layer MLP with a sigmoid activation :

\begin{equation}
  \begin{aligned}
    \boldsymbol{w_t} %&= [w^1_t, w^2_t,\dots, w^N_t] \\
      &= F_{as}(F_{fe}(x_t),\Theta) 
      %&= \sigma (g(F_{fe}(x_t),\Theta) )\\
      = \sigma(\Theta_2\delta(\Theta_1 F_{fe}(x_t))),
  \end{aligned}
\end{equation}
where $\delta$ represents the Tanh activation, $\Theta_1 \in \mathbb{R}^{K\times K} $ and $\Theta_2 \in \mathbb{R}^{N\times K} $. Then we rescale the weight vector  $\boldsymbol{w_t}$ and the similarity vector $\boldsymbol{\tilde{v}_t}$  as :

\begin{equation}
    u^i_t = F_{scale}(w^i_t,\tilde{v}^i_t) = w^i_t \tilde{v}^i_t, 
\end{equation}
where $\boldsymbol{u_t} = [u^1_t,u^2_t,\dots,u^N_t] $ is the input feature for attack model and $F_{scale}$ refers to a multiplication between the weight vector $\boldsymbol{w_t}$ and the similarity vector $\boldsymbol{\tilde{v}_t}$. 
%The pseudocode for adding the anchor select module is shown in Algorithm 2. 
We refer to SD-MI attack with \textit{anchor selector module} as $M_{AS+SD}$.

\section{Experimental Setup}
In this section, we introduce the configuration and implementation details of our experiments. 
\subsection{Datasets}
We use two datasets for Re-ID (Market1501 \cite{zheng2015scalable}, DukeMTMC-Re-ID \cite{zheng2017unlabeled}).  
The Market1501 contains 1501 different pedestrian classes with a total of 32,668 images from five high-resolution cameras and one low-resolution camera. The 751 pedestrian classes are used for the training set and the other for the test set (gallery set), and one image of each pedestrian in the test set is picked as a query to evaluate Re-ID model. On the other hand, the DukeMTMC-Re-ID has 16522 training images (from 702 pedestrians) and 17661 test images (gallery set) (from other 702 pedestrians) from eight static HD cameras at Duke University. Its query set is also chosen from the gallery set.
For  each dataset, we choose 2000 samples from the training set of Re-ID model and 2000 samples from the test set  to build the training dataset of attack model, 
and we also randomly sample 6000 images from the training set of Re-ID model  and  6000 images from  test set as the evaluation dataset for attack model.

\subsection{Target Models}
Our experiments select Re-ID target models with different backbone networks, including ResNet50 \cite{he2016deep}, MobileNetV2 \cite{sandler2018mobilenetv2} and Xception \cite{chollet2017xception}. 
We train these target models following the same setting as \cite{zhou2019omni,zhou2019torchreid,zhou2021learning}: 60 epochs, initial learning rate 0.0003, loss in Eq.\ref{formula1} and train batch size 32.

\begin{table*}[t]
\centering

\begin{tabular}{l|cc|cc|cc}
    \hline
    \hline
    \multirow{2}{*}{Method} & \multicolumn{2}{c|}{ResNet50} &\multicolumn{2}{c|}{MobileNetV2} &\multicolumn{2}{c}{Xception} \\
    \cline{2-7}
     &Market1501 & DukeMTMC &Market1501 & DukeMTMC &Market1501 & DukeMTMC\\
    \hline
    {$M_{FE}$} &80.1\% & 80.5\% & 74.9\%&72.7\% &78.5\% & 76.1\%\\
    {$M_{tloss}$} & 82.6\%& 86.2\% & 77.4\%& 77.8\% & 84.9\%&83.8\%\\
    {$M_{U\_low}$} &72.4\% & 70.8\% &65.5\% & 63.3\% &71.0\%  &66.1\%\\
    {$M_{U\_mid}$} &78.6\% & 77.4\% &71.0\% &69.3\% &76.9\% &72.3\%\\
    {$M_{U\_high}$} & 82.9\% &81.9\%  &74.0\% & 72.6\%& 79.6\% &75.9\%\\
    \hline    
    {$M_{SD}$ (ours)} &87.0\% & 88.7\%  & 80.6\%& 81.4\% & 89.7\% & 90.6\%\\
    {$M_{AS+SD}$ (ours)} & \textbf{87.3\%} & \textbf{89.1\%} &\textbf{81.2\%} & \textbf{82.2\%} &\textbf{90.1\%} &\textbf{91.6\%}\\
    \hline
    \hline
\end{tabular}
\caption{Performance comparison between the proposed method and existing membership inference attack baselines on different Re-ID models trained on Market1501 and DukeMTMC in terms of attack success  rate. The highest performance is marked in bold. }
\label{table1}
\end{table*}

\subsection{Baselines} 
\noindent \textbf{Feature based MI Attack} ($M_{FE}$). 
To verify the assumption that feature embedding does not characterize the train/test generalization gap as the direct model outputs do, we apply a feature embedding based MI attack method, which feeds the Re-ID feature of the target image into the same MI backbone as $M_{SD}$, following \citet{nasr2018comprehensive}. 

\noindent \textbf{Triple Loss based MI Attack} ($M_{tloss}$). 
This baseline follows the design of SOTA  black-box metric based MI attack \cite{sablayrolles2019white} that infers membership based on target image training loss and a hand-craft threshold. 
Since the cross-entropy based Re-ID losses is not directly accessible under our black box setting, we compute the  \textit{triple loss} \cite{schroff2015facenet} based on the image feature as a proxy loss.
Specifically, the triplet loss is obtained by setting the target image $x_t$ as the anchor image, and setting all the images with the same identity as positive samples and sampling $100$ images with other identities as negative samples. 

\noindent \textbf{User-level MI Attack} ($M_U$). 
We choose  user-level MI attack \cite{li2022user} as a comparison baseline, which is  designed for metric learning based models. 
This method can not be directly applied to instance-level MI attacks. As a result, we transfer the original method by sampling a set of images with the same identity as the target image and computing the intra-class distance based on the sampled images. 
To explore how the number of positive image in the identity affect  the performance of user-level MI attack, we report three results using the different number of sampled images. Specifically, $M_{U\_low}$, $M_{U\_mid}$ and $M_{U\_high}$ denotes the user-level MI attack sampling two, four and all  positive images for each target image respectively. 
Note that, this method requires the attacker to have the identity annotation of each pedestrian image and multiple  positive images for one identity, while our method does not have this requirement. 

\subsection{Evaluation Metrics}
Our experiments use attack success rate (ASR) \cite{shokri2017membership} as evaluation metrics, which is defined as the proportion of successful attacks (predicting members as members and non-members as non-members) to all unknown attacks. We also plot the curve of the comparison Receiver Operating Characteristic (ROC) to evaluate the trade-off between the true positive rate and false positive rate of the comparison methods. 

\section{Experiments}
We compare the MI attack performance of the proposed method with several baselines on Re-ID task. We also report the ablation study on the influences of different components and hyper-parameters on the performance of the proposed method. 
Finally, we show the performance comparison between our approaches and some SOTA methods on classification tasks.

\subsection{Performance Comparison}
Table \ref{table1} shows the ASR of our methods and the compared baselines attacking Re-ID models with different backbones (i.e. ResNet50, MobileNetV2 and Xception) trained on different datasets (i.e. Market1501 and DukeMTMC).
First of all, we observe that our approaches $M_{SD}$  significantly outperforms existing baseline methods in both datasets and all three Re-ID backbones, which verifies the effectiveness of leveraging the relative similarity between samples for membership inference. 
By adding an anchor selector, $M_{AS+SD}$ achieves the highest ASR, which shows the importance of selecting proper anchors for different images.

We also observe that $M_{FE}$ achieves lower ASR compared to other methods, which further verifies the assumption that  feature embedding contains more information irrelevant to training data and individual feature embedding is less informative on the training set membership compared to methods considering the inter-sample similarities. 
The user-level method $M_{U}$ also outperforms feature-based methods on Market1501, showing the importance of inter-sample relationships. 
However, this method only considers the correlation among positive samples, resulting in inferior performance compared to $M_{SD}$  and $M_{AS+SD}$. 
Furthermore, we observe that the performance of $M_U$ is severely affected by the number of positive samples in each class. Specifically, by observing the experiment results of $M_{U\_high}$, $M_{U\_mid}$, $M_{U\_low}$, we find that ASR decreases as the number of positive images in the identity decreases. In conclusion, compared to our method,  user-level MI attack has a more strict requirement on background knowledge of the target images, including the identity annotation as well as large number of positive images for each identity. 

\begin{figure*}[ht]
    \centering
    \subfigure[Market1501-ResNet50]{\includegraphics[width=0.5\columnwidth]{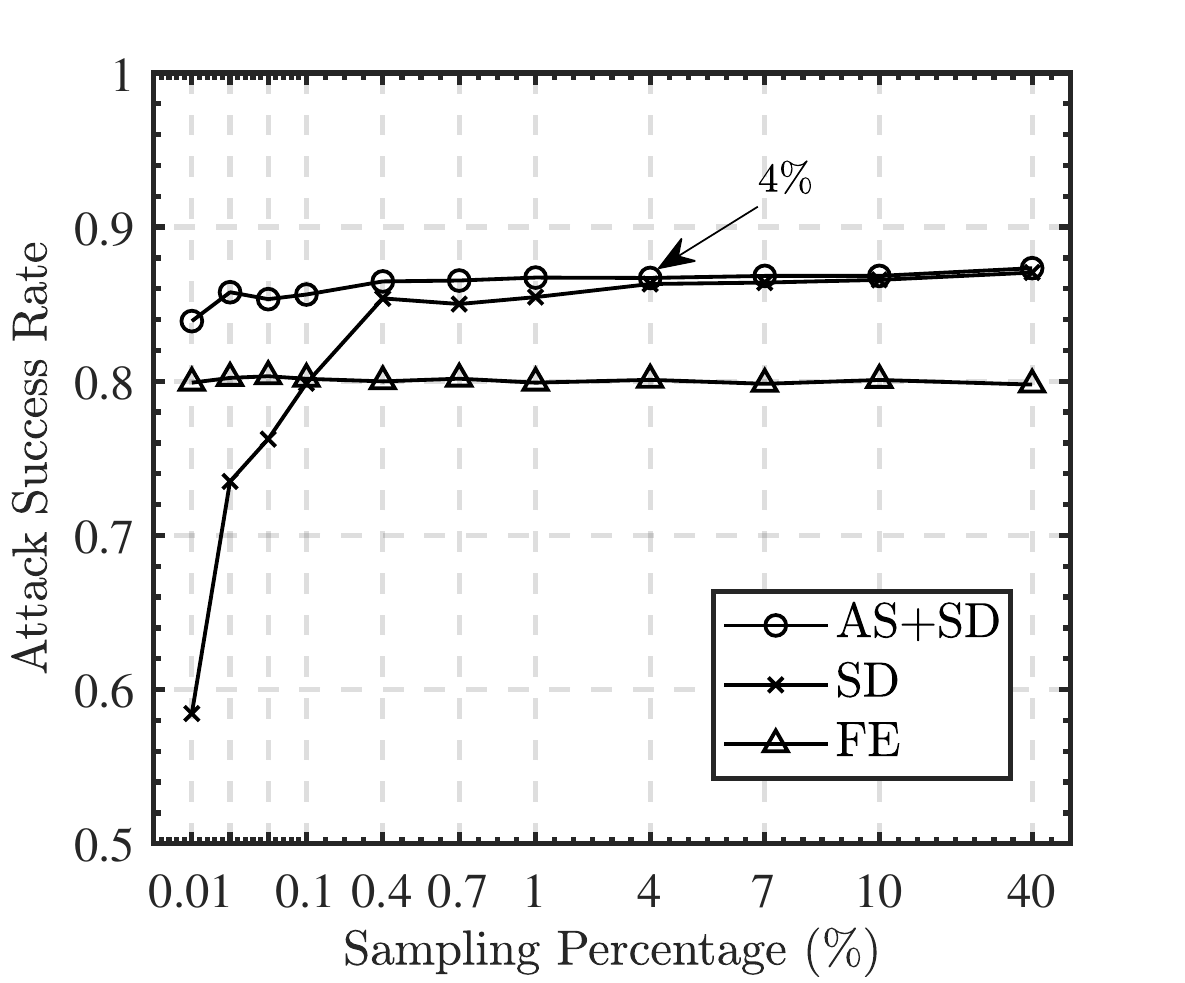}}
    \subfigure[Market1501-Xception]{\includegraphics[width=0.5\columnwidth]{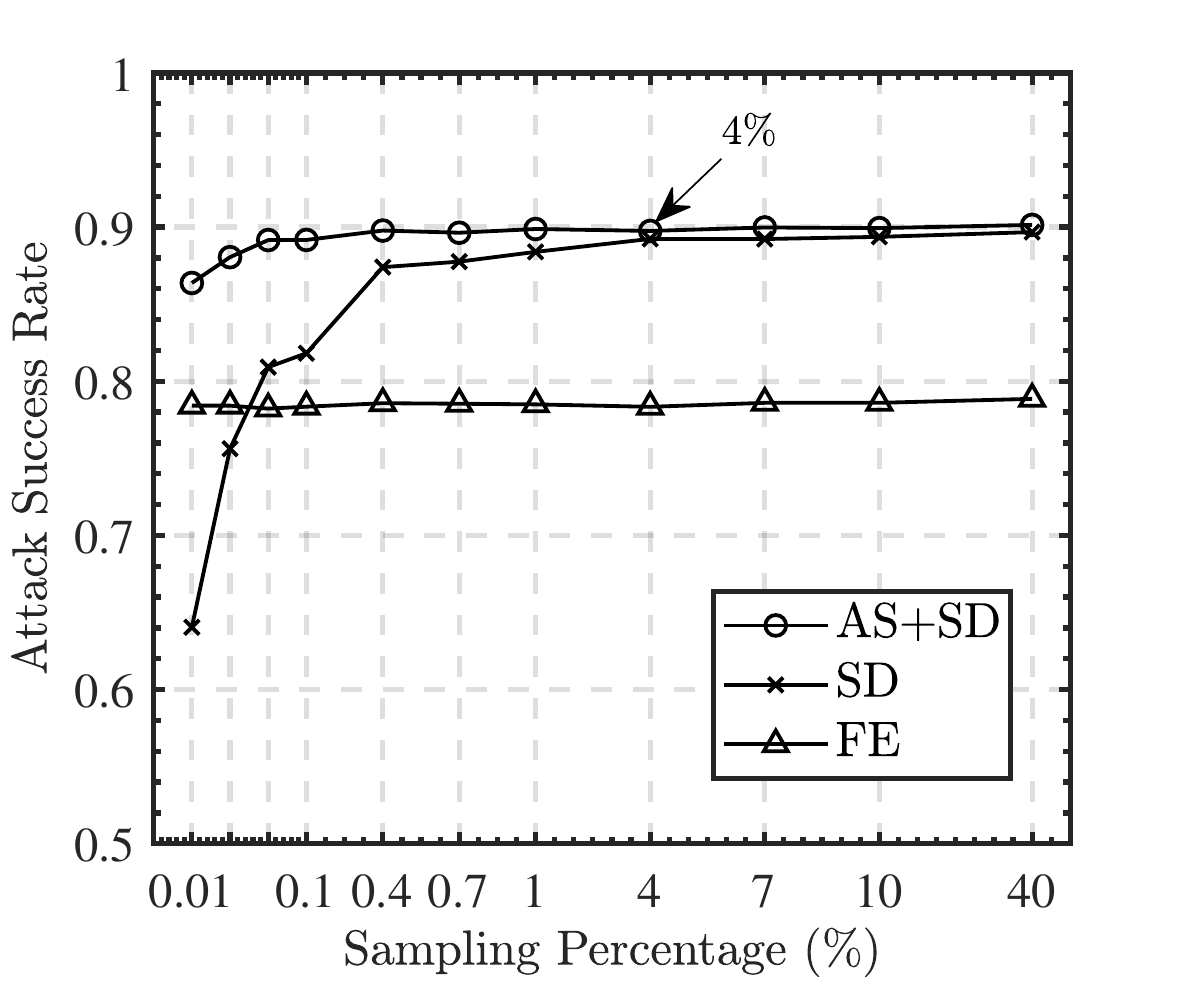}}
    \subfigure[DukeMTMC-ResNet50]{\includegraphics[width=0.5\columnwidth]{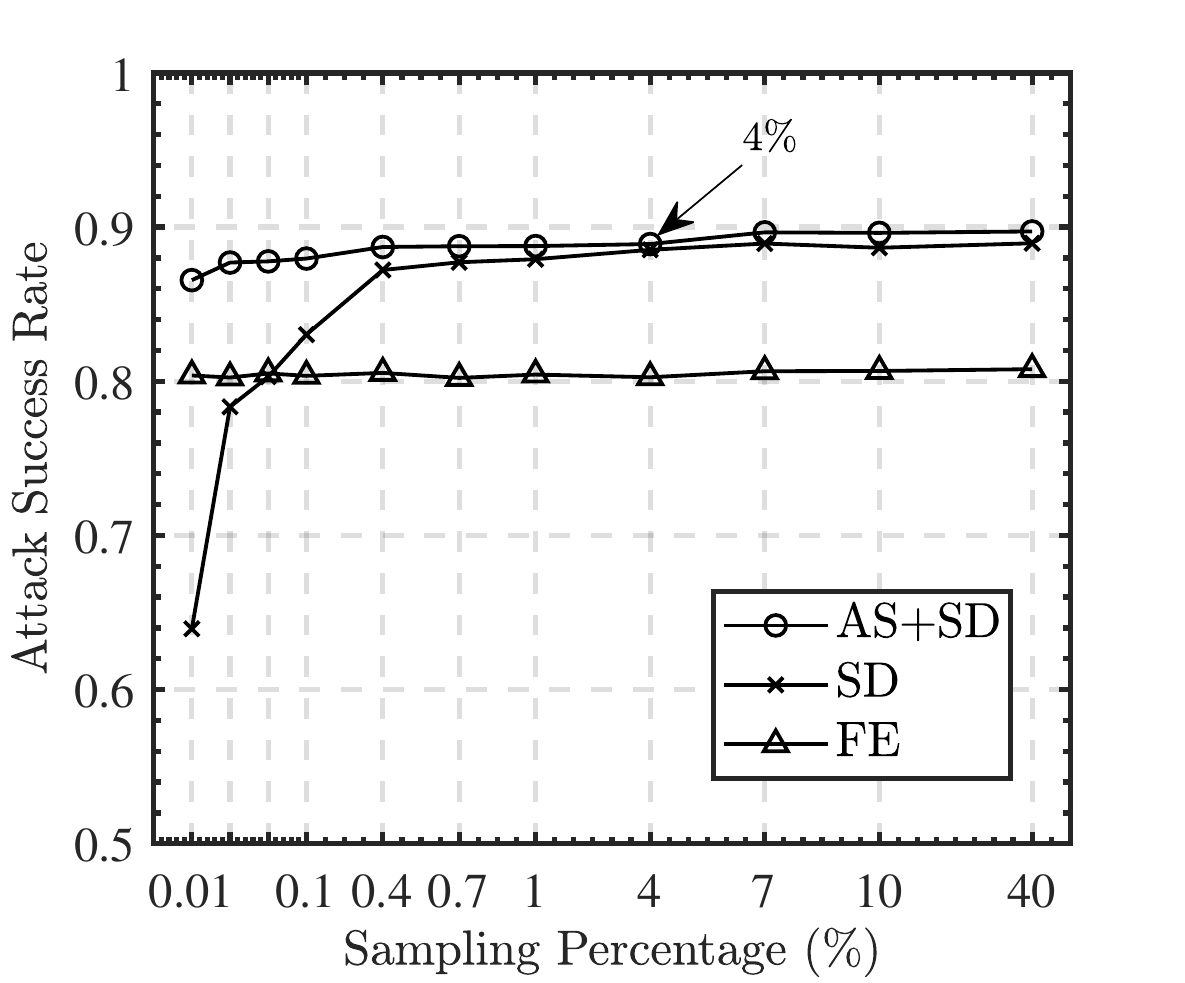}}
    \subfigure[DukeMTMC-Xception]{\includegraphics[width=0.5\columnwidth]{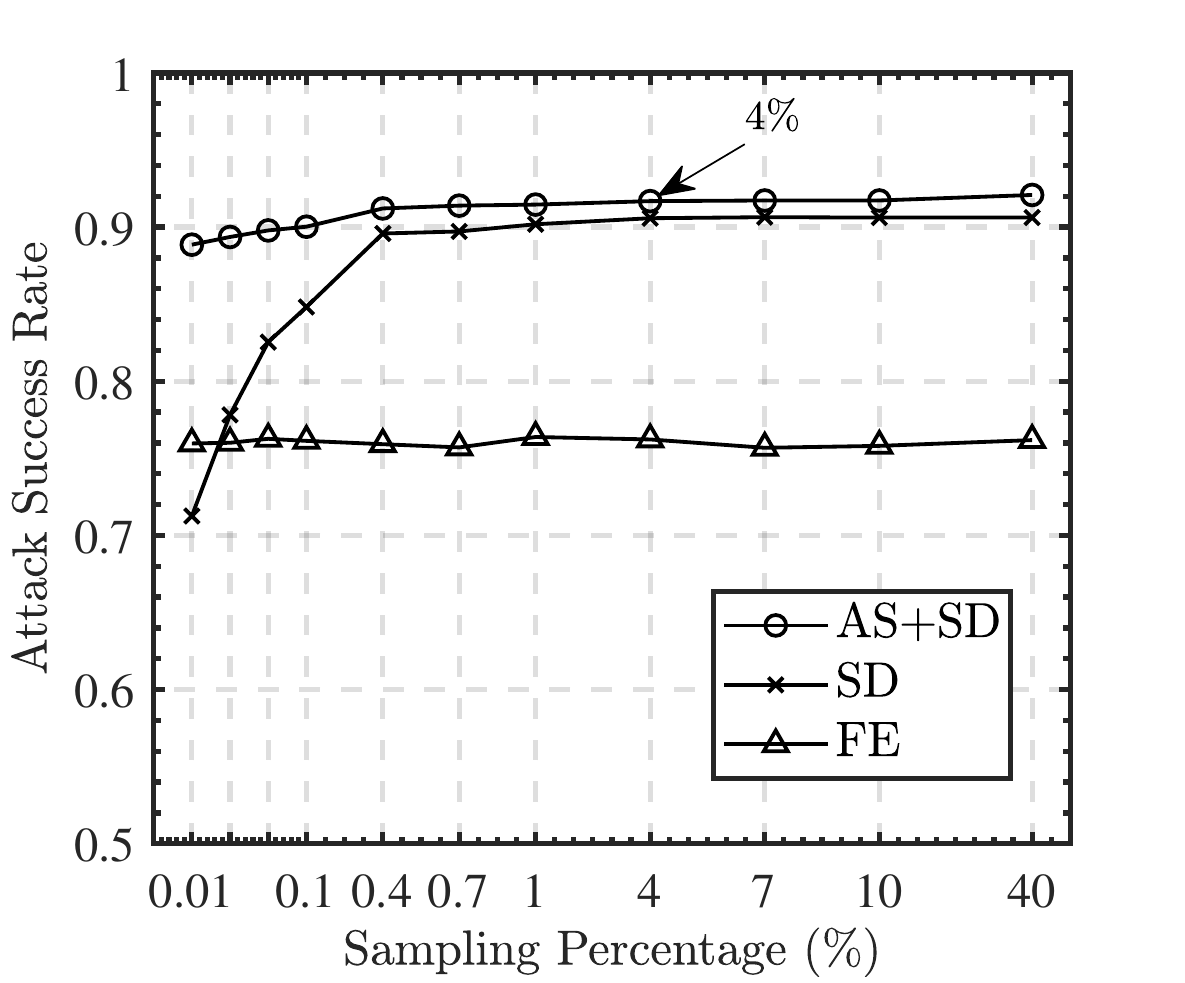}}
    \caption{Attack success rate with different sampling percentages of sampled anchor images in reference set for methods $M_{FE}$, $M_{SD}$ and $M_{AS+SD}$ on ResNet50 and Xception backbones trained on Market1501 and DukeMTMC datasets.}
    \label{fig7}
\end{figure*}

\begin{figure}
    \centering
    \includegraphics[width=0.85\columnwidth]{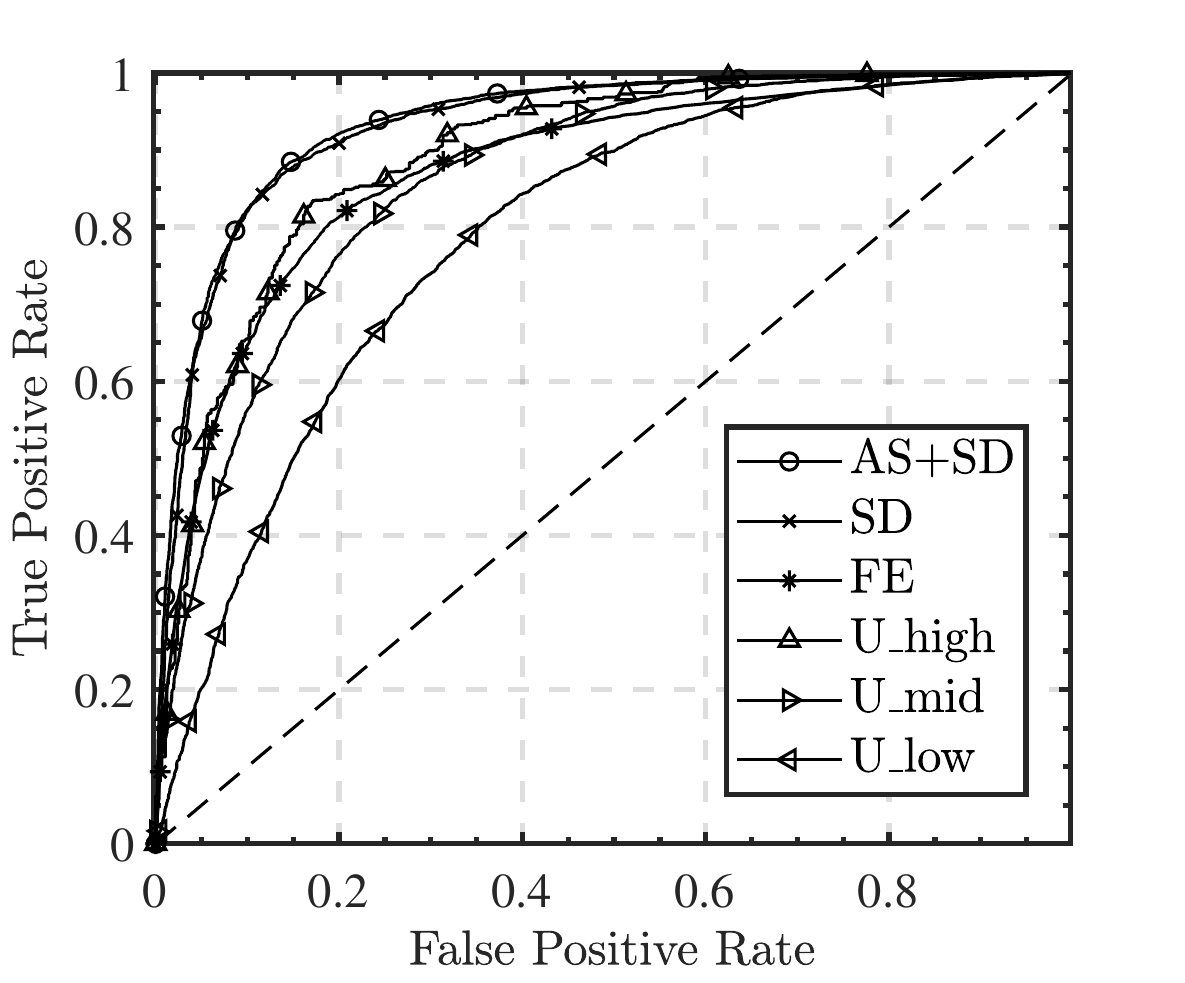}
    \caption{ROC curve of $M_{AS+SD}$, $M_{SD}$, $M_{U\_high}$, $M_{U\_mid}$, $M_{U\_low}$ and $M_{FE}$ on ResNet50 trained on Market1501. }
    \label{ROC}
\end{figure}

 We further compare our methods and the compared baselines on ResNet50 trained with Market1501 in terms of the ROC curve, as shown in Figure \ref{ROC}, where our methods $M_{AS+SD}$ and $M_{SD}$ outperform other  methods with the  highest  \textit{Area Under Curve (AUC)} values 0.935 and 0.930.
 
\subsection{Reference Set Sampling }
Based on the formal analysis in Section 3, we find that it is essential to select proper reference images as proxy center to approximate the learned identity centers. Intuitively, if large enough reference set is used, we can always find samples which is close enough to the identity center. However, when the reference set is small, there may not be enough samples to properly approximate the identity centers. 
As a result, Figure \ref{fig7} shows how the percentage of sampled anchor images in the reference set affects the  attacking successful of $M_{SD}$ and $M_{AS+SD}$. 
We observe that ASR of $M_{SD}$ drastically decreases as the percentage of the sampled reference images decreases. 
On the other hand, by adding an extra anchor selector to assign higher importance weight to proper anchor images, $M_{AS+SD}$ significantly outperforms $M_{SD}$ when the number of anchors is low and achieves performance upper-bound  when only $4\%$ of the images are sampled, which verifies the importance of selecting proper reference image to approximate the identity anchors. 

\subsection{Evaluation on Classification}
To examine how our proposed method works for tasks other than Re-ID, we apply the proposed method to classification task and compare its performance with several state-of-the-art MI attack methods. 
We select  CIFAR10 \cite{krizhevsky2009learning} as our benchmark dataset with the target model  ResNet18 \cite{he2016deep}, ResNet50 \cite{he2016deep}, VGG19 \cite{simonyan2014very} and GoogLeNet \cite{szegedy2015going}. 
The target models are trained with SGD optimizer with a learning rate 0.1, 200 epochs and $l_2$ regularization with weight  set to 0.0005. The comparison methods include logits-based MI attack  $M_{logits}$  \cite{shokri2017membership,salem2018ml}  that feeds the output logits into the attack neural network, feature-based method $M_{FE}$ and loss-based MI attack $M_{loss}$ \cite{sablayrolles2019white} which conducts MI based on the classification loss and a manually defined threshold. 
As shown in Table \ref{table2}, our algorithm $M_{AS+SD}$  achieves comparable ASR to previous SOTA  algorithms $M_{loss}$ on most target models, and the higher ASR on ResNet18, which shows that the inter-sample similarity  also contains the sufficient information about generalization gap between training and test set in the classification task. 

\begin{table}[t]
\centering
\begin{tabular}{c|cccc}
    \hline
    \hline
     Model &   $M_{loss}$ & $M_{logits}$ &  $M_{FE}$ & $M_{AS+SD}$ \\
    \hline

    {ResNet18} & 78.7\%& 78.5\% & 78.6\%& 79.0\%\\

    {ResNet50} & 69.1\% &67.9\%&66.9\% &  68.3\% \\

    {VGG19} &   63.9\% & 63.8\%& 63.6\%& 63.6\%\\

    {GoogLeNet} &  67.8\% &65.9\% &63.9\% & 66.9\%    \\
    \hline
    \hline

\end{tabular}
\caption{Performance comparison between the proposed method and existing membership inference attack baselines on different classification models trained on CIFAR10 in terms of attack success rate. }
\label{table2}
\end{table}

\section{Conclusion}
This paper raises a rarely studied privacy risk of the training data of person re-identification. The information leakage from Re-ID data can be quantified by membership inference attack. 
However, Re-ID is a fine-grained recognition task with complex feature embedding, and model outputs commonly used by existing MI like logits and losses are not accessible during inference. 
As a result, this paper conducts both formal and empirical analysis to discover a new set of feature for Re-ID MI attacks, which is the inter-sample similarity of image pairs.
As a result, a novel membership inference attack method is proposed to quantify the information leakage of the Re-ID dataset by exploiting the inter-sample correlation between pedestrian images. The proposed method outperforms existing MI attack approaches on Re-ID models. 

\section{Acknowledgments}
This work was supported by National Natural Science Fund of China (62076184, 61976158, 61976160, 62076182), in part by Shanghai Innovation Action Project of Science and Technology (20511100700) and Shanghai Natural Science Foundation (22ZR1466700), in part by   Fundamental Research Funds for the Central Universities and State Key Laboratory of Integrated Services Networks (Xidian University).

\bibliography{aaai23}

\end{document}